\documentclass{article}
\usepackage{graphicx}
\usepackage{verbatim}
\usepackage{authblk}
\title{Universal rapid machine learning models for predicting unconvoluted and convoluted X-ray Absorption Spectra}
\author[1]{Fei Zhan}
\author[1]{Zhi Geng}
\affil[1]{Institute of high energy physics,Chinese academy of sciences}

\usepackage{chemformula} 
\usepackage[T1]{fontenc} 
\usepackage{graphicx}
\usepackage{verbatim}
\usepackage{float}
\usepackage{tabularx}
\usepackage{booktabs}
\usepackage{changepage} 
\usepackage{multirow}
\newlength{\widthnormal}
\newlength{\widthwide}
\newlength{\leftcolumnlength}
\newlength{\extralength}
\newlength{\fulllength}
\begin{document}
\maketitle

\begin{abstract}

\quad 
X-ray absorption near edge structure (XANES) is an essential tool for elucidating the atomic-scale, local three-dimensional (3D) structure of given materials and molecules. The rapid computation of XANES based on molecular 3D structures constitutes a vital element of quantitative XANES analysis.
Here, we present an XANES prediction model. It takes 3D structures as input and generates either unconvoluted XANES or convoluted spectra as output, demonstrating excellent generalizability across diverse instrumental broadening. This model has validated its predictive capability for both hard X-ray XAS (exemplified by K-edges of 3d 4d metals and lanthanides) and soft X-ray XAS (using S K-edge as examples). Adopting the model, XANES spectra of multiple elements can be predicted using a single unified model. A highly efficient 3D structure fitting algorithm based on this unconvoluted XANES prediction model, aiming to serve as an online data analysis method suitable for XAS beamlines.

\end{abstract}	
\section{Introduction} 
X-ray absorption spectroscopy (XAS) stands as a highly potent technique for probing the local atomic structure surrounding the absorber in diverse materials. In light of the imperative to rapidly characterize materials based on spectral data, there has been a surging interest in devising efficient methodologies for XAS analysis.
The capability of machine learning to autonomously discern patterns and correlations within vast datasets, combined with its minimal time requirements and decreased reliance on specialized user knowledge, positions it as an optimal choice for accelerated data analysis in the realm of XAS.
Since 2017, machine learning models have enabled the extraction of material properties from spectra, with reported success in predicting metrics such as coordination number\cite{timoshenko2017JPCL,MLCNJPCL2020}, bond length\cite{MLbindlenghthnpj2021},oxidation state\cite{MLoxidationstatenpj2025,MLoxidationstatenpj2024}, radial distribution function\cite{timoshenko2018PRL,timoshenko2018RDFNanoLetters}, and G2 terms of wACSF\cite{PenfoldG2wascf}.
Among material properties, 3D structures hold paramount importance as they are instrumental in elucidating microscopic mechanisms underlying strain effects, chemical reactions, and structure–function relationships. Nevertheless, existing machine learning approaches remain largely unable to conduct direct quantitative analysis of 3D structures based on XAS data.

\par
The establishment of an efficient and accurate machine learning framework for direct prediction of XANES spectra from 3D structures remains a critical objective in computational materials science.
Graph neural network (GNN), a machine learning model used a lot in chemistry such as prediction of potential energy and kinetic energy in Ab initio molecular dynamics simulations\cite{GNNJPCLmaji2025}, prediction of atomization energy and excited state energy
\cite{GNNJCASli2024}, prediction of protein binding afffnityetc\cite{GNNJCIMzhou2024} etc.
Capacity of GNN to represent 3D structure through graph topologies (with atoms as nodes and bonds as edges) makes it particularly well-suited for capturing complex structural relationships.
It have been applied to predict N and O K-edge XAS by Carbone\cite{carbone2020prl}, and to C K-edge XAS by Bande and Kotobi et al\cite{GNNJACS}. These works all involve inputting 3D structures of small organic molecules for soft X-ray XAS predictions.
The extension of application scope and improvement in accuracy for predicting X-ray absorption spectroscopy (XAS) through inputting 3D structures represents a continuously evolving research direction that warrants in-depth investigation.
Among available GNN models, 3D GNNs representing atoms as nodes and interatomic bonds as edges, represent suitable framework for analyzing 3D structures. Prominent 3D GNN variants including SchNet\cite{schnet2018}, SphereNet\cite{spherenet2021}, and DimeNet++\cite{dimenet2020} differ fundamentally in two key aspects: the computational efficiency and comprehensiveness of their geometric features and methodologies for characteristic feature functions. Based on 3D GNN model, we present our customized XANES prediction model. It takes 3D structures as input and generates either unconvoluted XANES or convoluted spectra as output, demonstrating excellent generalizability across diverse instrumental broadening. Adopting the model, XANES spectra of multiple elements can be predicted using a single unified model. A highly efficient 3D structure fitting algorithm based on this unconvoluted XANES prediction model, aiming to serve as an online data analysis method suitable for XAS beamlines.

\section{Method}

As mentioned above, 3D GNNs represent a class of models well suited for predicting XAS from 3D structures.
Subsequently, we propose XAS3D which is a variant of 3D GNN specifically adapted for XAS prediction.
This model takes 3D structure as input and generates the predicted spectrum as output.
The input 3D structure is represented by a cluster centered on the absorbing atom, a representation aligned with XAS characteristics, which aims to capture information about the local atomic environment surrounding the absorbing atom.
In a 3D GNN model, the input 3D structure is divided into two components, atomic species and Cartesian coordinates of atoms.
Firstly, within 3D GNN model, embedding layers transform discrete atomic species into a continuous numerical vector v (the initial node feature vector), which is subsequently updated iteratively through graph neural network convolutional layers.
Secondly as another component of the input, Cartesian coordinates of atoms are first processed to derive geometric features, including distances r, angles $\theta$, and dihedral angles $\phi$ for atom pairs (e.g., edges in the graph). Following the approach developed by Limei Wang\cite{comenet2022}, we calculated geometric features using their established framework. Subsequently, feature function values are computed based on these precomputed geometric features, which serve as edge weights in the graph convolution layers. Finally, the atomic node feature vector v, after passing through several weighted graph convolutional layers, is pooled to obtain the spectra corresponding to the entire graph.
Here, we propose XAS3D a refined 3D GNN architecture retaining exclusively those edges connected to the absorbing atom. This strategic edge filtering mechanism suppresses non-essential information while preserving critical local coordination environments.
This model demonstrates predictive capability for both hard X-ray and soft X-ray XANES, enabling multi-element XANES prediction via a single unified model. Reliable results are attainable even with scarce 3D structural data for specific elements.
Furthermore, the model can extend its predictive capability from conventional convoluted spectra to unconvoluted XANES, thereby enhancing its applicability across diverse spectroscopic analyses.

\par
As a crucial component of model training and validation, here we briefly introduce the dataset.
A dataset linking three-dimensional structures with corresponding X-ray absorption spectrum was developed to train and validate machine learning models.
The three-dimensional local structures associated with XAS originate from crystallographic structural data deposited in the Cambridge Crystallographic Data Centre (CCDC).
Crystallographic structural data derived from X-ray or neutron diffraction experiments represent stable structural information and therefore do not require secondary stability assessment using metrics like $E_{hull}$\cite{Ehull0.25} (as necessitated for structures in the Material Project database). For duplicate chemical formulas, only one structure was retained. This ensured that structures in the training and validation sets corresponded to distinct chemical formulas, preventing samples with identical formulas from appearing simultaneously in both sets—a design choice critical for unbiased model evaluation.
FDMNES, one of the most popular XAS simulation package, is used to perform XAS simulation in multiple scattering mode. 
Employing multiple scattering theory, we adopted the Real Hedin-Lundqvist parametrization of the local density approximation (LDA) to model the exchange-correlation potential. A cluster extending to a radius of 5 \AA centered on the absorbing atom sufficed to ensure convergence within the self-consistent cycle coupling charge density, Dyson's equation, and Green's functions\cite{joly2009self}.
Thus, a "3D structure-XAS" dataset was established. For each dataset, 80$\%$ was allocated to the training set, 10$\%$ to the validation set, and 10$\%$ to the test set.
To facilitate dataset visualization, given element-related datasets were clustered into 10 categories using the k-means algorithm to illustrate the spectral variation range covered by the dataset. For the 3d transition metal dataset, where all elements were treated as a single integrated dataset during training and validation, individual spectral clustering analyses were performed separately for each of the ten elements from Sc to Zn. It can be observed that there is a significant variation in the spectral shape, which confirms that the dataset contains a diverse range of structural changes as illustrated in Support Information.
Although trained with economically efficient FDMNES multiple scattering calculated data, our method exhibits natural adaptability for extending to spectrum generated across different theoretical methods, such as finite difference calculations\cite{FDMJCTC}, Bethe-Salpeter equation method\cite{vinson2011BSE}, TDDFT\cite{george2008TDDFT} and so on.

\section{Hyperparameters optimization of machine learning models}
\begin{figure}[H]
	\centering
	\includegraphics[width=0.7\linewidth]{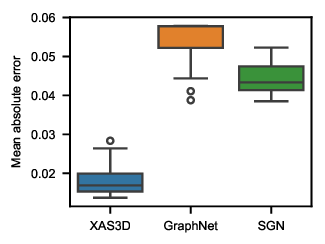}
	\caption[short]{Boxplots of performance metric (MAE) of XAS3D, GraphNet and SGN GNN models using different hyperparameters in Ni K-edge XANES prediction.}
	\label{fig:SIBoxHyper}
\end{figure}

To prevent performance degradation of the models due to improper hyperparameter selection, 72 sets of hyperparameters were searched for each model. 
Specifically, a grid search was performed on three shared hyperparameters across all three GNN models: the number of layers (nl), the hidden embedding size (hidden) per GNN layer, and the learning rate (lr).
Six values were tested for the number of layers (1, 2, 3, 4, 6, 8), six for the hidden embedding size (16, 32, 64, 96, 128, 256), and two for the learning rate (0.001, 0.0005). Through grid sampling across these ranges, a total of 72 paramter configurations were generated (6 × 6 × 2).
Each GNN model's unique hyperparameters were sampled in 72 sets, which were then combined with the shared hyperparameters to form 72 complete hyperparameter configurations for the given model.
For instance, in the SGN model: The K value hyperparameter adopted nine discrete integers from 2 to 10 inclusive. By repeating this sequence eight times and applying random permutation, we obtained 72 distinct K value instances. When integrated with previously sampled general GNN hyperparameters, these yielded 72 comprehensive hyperparameter configurations for SGN.
Analogously, the XAS3D model’s unique hyperparameters—including Distance embedding dim, Angle embedding dim, Middle embedding size, and Output block layer count—underwent independent sampling to produce 72 variants each. Integration with the with previously sampled general GNN hyperparameters resulted in 72 comprehensive hyperparameter configurations for XAS3D.
We performed hyperparameter optimization for the XAS3D, GraphNet, and SGN models on the Ni XANES prediction dataset.
The performance across 72 hyperparameter sets for these three models is presented in Fig.\ref{fig:SIBoxHyper}.
The performance boxplots across 72 hyperparameter sets for these three models are shown in Fig.\ref{fig:SIBoxHyper}.
Among the 72 sets of hyperparameters evaluated, XAS3D demonstrated a median value of 0.0169, outperforming both GraphNet (0.0578) and SGN (0.0433), indicating its overall superior performance.
Under their respective optimal hyperparameter configurations, the models attained the following MAE scores on the testing set: 0.0137 (XAS3D), 0.0388 (GraphNet), and 0.0385 (SGN). On the validation set, the corresponding MAE metrics were 0.0128 (XAS3D), 0.0389 (GraphNet) and 0.0397 (SGN).

\section{Results $\&$ Discussion} 

\begin{figure}[H]
	\centering
	\includegraphics[width=\widthnormal]{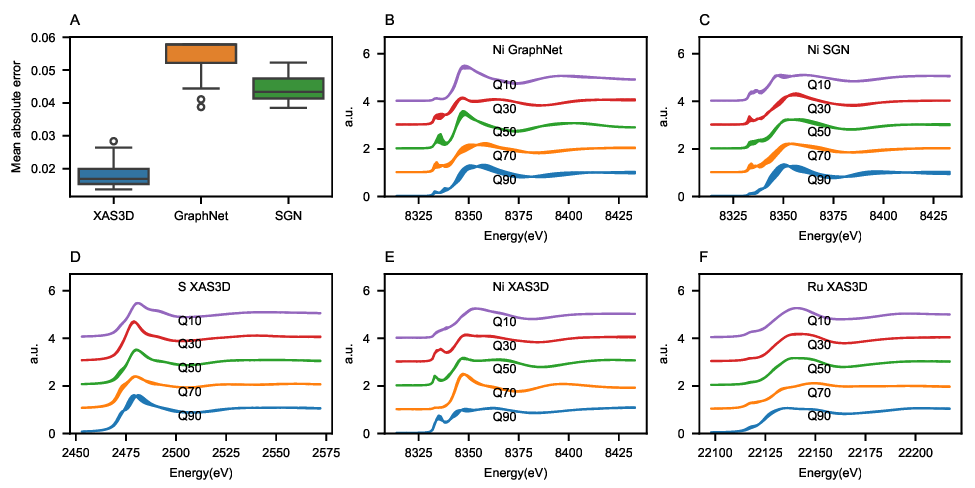}
	\caption[short]{(\textbf{A}):Boxplots of performance metric (MAE) of XAS3D, GraphNet and SGN GNN models using different hyperparameters in Ni K-edge XANES prediction. (\textbf{B}):Prediction results generated by the GraphNet model for Ni K-edge XANES prediction at different quantile levels from q10 through q90. (\textbf{C}):Prediction results generated by the SGN model for Ni K-edge XANES prediction at different quantile levels from q10 through q90.(\textbf{D}):Prediction results generated by the XAS3D model for S K-edge XANES prediction at different quantile levels from q10 through q90. (\textbf{E}):Prediction results generated by the XAS3D model for Ni K-edge XANES prediction at different quantile levels from q10 through q90. (\textbf{F}):Prediction results generated by the XAS3D model for Ru K-edge XANES prediction at different quantile levels from q10 through q90. In the above subplots, XANES simulated with multiple scattering framework(solid lines) and with machine learning model(dashed lines) on validation dataset.}
	\label{fig:singlemodel}
\end{figure}

In practice, an XANES prediction model was constructed for individual elements, exemplified here using the K-edge XANES of Ni, S and Ru. Taking Ni as a representative case to evaluate model performance, three GNN architectures XAS3D, GraphNet\cite{graphnet2019}, and SGN\cite{SGN2019} were constructed for XANES prediction. To prevent performance degradation of the models due to improper hyperparameter selection, 72 sets of hyperparameters were searched for each model. For further details, see Support information section "Hyperparameters optimization".
Specifically, three shared hyperparameters across all three models the number of layers in the model, the hidden channels per GNN layer, and the learning rate were sampled using an identical grid of 72 points. Meanwhile, other model-specific hyperparameters for each model were sampled via a hypercube sampling method.
Fig.\ref{fig:singlemodel} illustrates the distribution of MAEs across these configurations.
Among the 72 sets of hyperparameters evaluated, XAS3D demonstrated a median value of 0.0169, outperforming both GraphNet (0.0578) and SGN (0.0433), indicating its overall superior performance.
Under their respective optimal hyperparameter settings, the models achieved mean absolute errors (MAE) as follows: 0.0128 for XAS3D, 0.0389 for GraphNet, and 0.0397 for SGN.
To comprehensively visualize performance variability across models, we ranked them according to their validation set metrics and plotted spectral predictions covering five quantiles (q10–q90), as illustrated in Fig.\ref{fig:singlemodel}.
For Ni/XAS3D model: predictions showed visually indistinguishable agreement with theoretical spectra up to q70; subtle deviations emerged only at q90. In contrast, both GraphNet and SGN exhibited discernible discrepancies even at q10, with progressive divergence through q30–q70, reaching substantial mismatches in spectral trends by q90. 
The S/XAS3D model registered an MAE of 0.0197, higher than Ni/XAS3D model 0.0130, likely due to inherent challenges in capturing electronic structure effects without explicit modeling. As a soft X-ray probe (absorption edge at 2472 eV), sulfur XAS contains rich electronic information derivable solely from geometric inputs. 
In terms of spectral prediction performance, excellent agreement between predicted and theoretical spectra was observed up to the q50 quantile. However, the predictive accuracy began to deteriorate starting from the q70 quantile.
At the q90 quantile, the predicted position of the white line peak within the range of 2475–2485 eV exhibits an energy shift towards higher values by 2.3 eV.
For ruthenium representative of 4d transition metals or platinum-group elements, the Ru/XAS3D model achieved an exceptionally low MAE of 0.0082. Even at q90, predicted spectra aligned closely with theoretical one.
Considering the core-hole width of the K-edge of Ru (5.33 eV) is significantly larger than that of Ni (1.44 eV) and S (0.59 eV).
Besides the model's favorable performance, this result is also attributable to the significant broadening of Ru element core-hole excitations, which obscures fine structural details of cross-section in theoretical XANES.
Building upon the good performance of single-element models, we conducted studies on multi-element models.
Motivated by the above single-element performance, we next asked whether multi-element XANES prediction models possess potential advantages beyond their inherent generalization capability.
Our findings demonstrate that multi-element models utilize knowledge acquired from other elements to improve the predictive performance for elements with limited data availability. 

\par
In practical applications, not every element has a large number of samples for the given absorption edge XAS to conduct independent model training. To simulate this scenario, we conducted a test with Ni as the research element, including 50-8000 samples. For comparison, baseline models were trained solely on the Ni element, with both the training and validation sets comprising only Ni element.
To investigate the influence of samples from other elements in the training set during multi-element XAS prediction model training, besides Ni, 34,000 samples of four other 3d or 4d transition metal elements were included for model training. The validation set contained no Ni samples, consisting only of samples from other 3d or 4d transition metal elements. This design was implemented because, in practice, when samples of a specific element (e.g., Ni) are scarce, it is desirable to include as many of its samples as possible in the training set, leaving insufficient samples for the validation set. Consequently, this study evaluated the predictive performance of models optimized using non-Ni element validation sets on Ni XANES spectra, results are shown in Tab.\ref{tbl:MAE3d4d}. It can be observed that when the dataset includes 34,000 samples of Fe, Co, Cu, and Zn elements, even with only 50 Ni samples included, the model can achieve a satisfactory prediction performance with an MAE of 0.0175. 
It is evident that samples from other 3d transition metal elements significantly enhance the prediction accuracy for Ni XANES spectra. Moreover, the enhancement ratio increased as the number of available Ni samples decreases. For instance, with only 50 Ni samples, incorporating 34,000 samples each of Fe, Co, Cu, and Zn into the training set reduced the model's Mean Absolute Error (MAE) by 80\% compared to training solely on Ni samples. This MAE reduction percentage decreased to 30 percentage points when the Ni sample count was 500, and further to 14\% at counts of 2,000 or 4,000. When the Ni sample size was sufficiently large (8,000), adding extra 3d transition metal samples provided no additional improvement in model performance. In contrast, incorporating 4d transition metal samples showed negligible impact on Ni XANES prediction performance, except under conditions of extreme Ni scarcity (e.g., 50 samples).

\par
\begin{table} 
	\caption{Performance metrics MAE in the testing dataset for XANES prediction of Ni element. MAE 3d/ MAE 4d denotes that the model training set comprises samples of Ni element and 34,000 samples of 3d/4d transition metals. MAE Ni indicates that the training set contains only Ni element samples.}	
		\begin{tabularx}{\linewidth}{lccc}
			\toprule
			Number of Ni samples 
			\\in training set  &MAE 3d   & MAE 4d   &MAE Ni \\
			\midrule
			50                &0.0175    &0.0301     &0.0869  \\
			\midrule
			500               &0.0156    &0.0244     &0.0223 \\
			\midrule
			2000              &0.0145    &0.0165     &0.0169  \\
			\midrule
			4000              &0.0128    &0.0146     &0.0149 \\
			\midrule
			8000              &0.0130    &0.0133     &0.0129\\
			\bottomrule
		\end{tabularx}
	\label{tbl:MAE3d4d}
\end{table}

\par
Building upon the good performance of single-element models, we conducted studies on multi-element models. We will discuss the unconvoluted XANES prediction model for multiple elements. This model has practical application scenarios. In experiments, different beamlines have different instrumental broadening. Even for second- and third-generation synchrotron radiation sources, variations in the degree of optimization of the white beam mirror posture can lead to differences in the light divergence on the first crystal of the monochromator, resulting in changes in the energy resolution of the beamline and, consequently, variations in the experimental spectral broadening. In terms of data processing workflows, software such as MXAN and Pyfitit requires continuous adjustment of the XANES broadening parameters during the fitting process to match the experimental spectra. Replacing the multiple scattering calculations with a model that predicts unconvoluted XANES can improve data analysis efficiency. However, models that predict XANES with fixed broadening are limited. Based on the needs of experiments and data analysis, we have constructed an unconvoluted XANES prediction model covering all 3d metal elements from Sc to Zn. The prediction results are shown in the figure. It can be seen that the model can predict the main features and spectral shape trends of the unconvoluted XANES, including the position of the pre-edge peak, although the intensity prediction of the pre-edge peak is less accurate. To compare with the convoluted XANES prediction model, the unconvoluted XANES prediction results were broadened using the same broadening parameters and the MAE was calculated, as shown in Figures 1(A) and (E)-(G). After convolution, the performance of the unconvoluted XANES prediction results is generally slightly worse than that of the convoluted XANES model. However, from the perspective of spectral line prediction, it still meets the requirements for quantitative analysis. For example, for elements like Sc and Fe, the performance of the convoluted unconvoluted XANES prediction is mainly poor in terms of the intensity of the pre-edge peak at the q90 quantile, but the prediction of the pre-edge peak position is good. In summary, the unconvoluted XANES prediction model, with only a minor loss in prediction performance ranging from 4.4 percent(Ni) to 14.4 percent(Zn), can be flexibly applied to data analysis under different beamlines and experimental conditions, and it also allows for the fitting of broadening parameters.

\begin{figure}[H]
	\centering
	\includegraphics[width=\widthnormal]{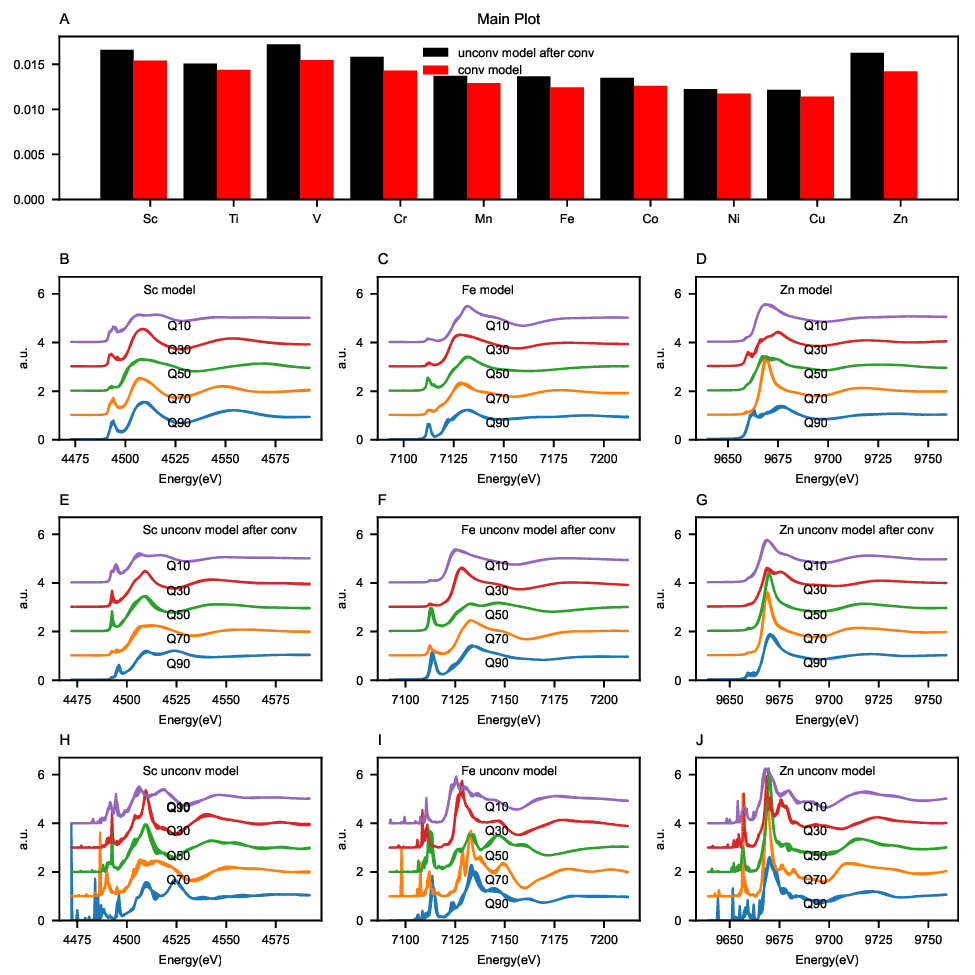}
	\caption[short]{(\textbf{A}):MAE of convoluted XANES predictions for 3d transition metals versus MAE calculated after applying convolution to results of unconvoluted XANES prediction models. (\textbf{B}):Prediction results generated by the XAS3D model for Sc K-edge XANES prediction at different quantile levels from q10 through q90. (\textbf{C}):Prediction results generated by the XAS3D model for Fe K-edge XANES prediction at different quantile levels from q10 through q90. (\textbf{D}):Prediction results generated by the XAS3D model for Zn K-edge XANES prediction at different quantile levels from q10 through q90. (\textbf{E}):The XAS3D model first predicted unconvoluted Sc K-edge XANES, then applies convolution to unconvoluted spectrum. MAE were calculated between predicted and theoretical spectra both after convolution. The performance across q10 through q90 quantiles was presented. (\textbf{F}):The XAS3D model first predicted unconvoluted Fe K-edge XANES, then applies convolution to unconvoluted spectrum. MAE were calculated between predicted and theoretical spectra both after convolution. The performance across q10 through q90 quantiles was presented. (\textbf{G}):The XAS3D model first predicted unconvoluted Zn K-edge XANES, then applies convolution to unconvoluted spectrum. MAE were calculated between predicted and theoretical spectra both after convolution. The performance across q10 through q90 quantiles was presented. In the above subplots, XANES simulated with multiple scattering framework(solid lines) and with machine learning model(dashed lines) on validation dataset.}
	\label{fig:unconv3dmodelMAEBAR}
\end{figure}

\par
In the current discussions of published results related to the "structure to spectrum" model, there is limited attention given to how it can be applied for quantitative structural analysis of experimental spectra. Here we discussed it detailed in the Supporting Information section titled "Application in Experimental XAS." The simplest application of such models is the rapid calculation of spectrum for materials using given three-dimensional structures.
We computed spectrum for several standard samples using the XAS3D model and compared them with experimental ones, which can be seen in Support Information. The results demonstrate that the predicted spectra from the XAS3D model successfully reconstruct the primary characteristic peaks of the experimental spectra, exhibiting good agreement in both peak positions and relative intensities. 
\par
\begin{figure}[H]
	\centering
	\includegraphics[width=\widthnormal]{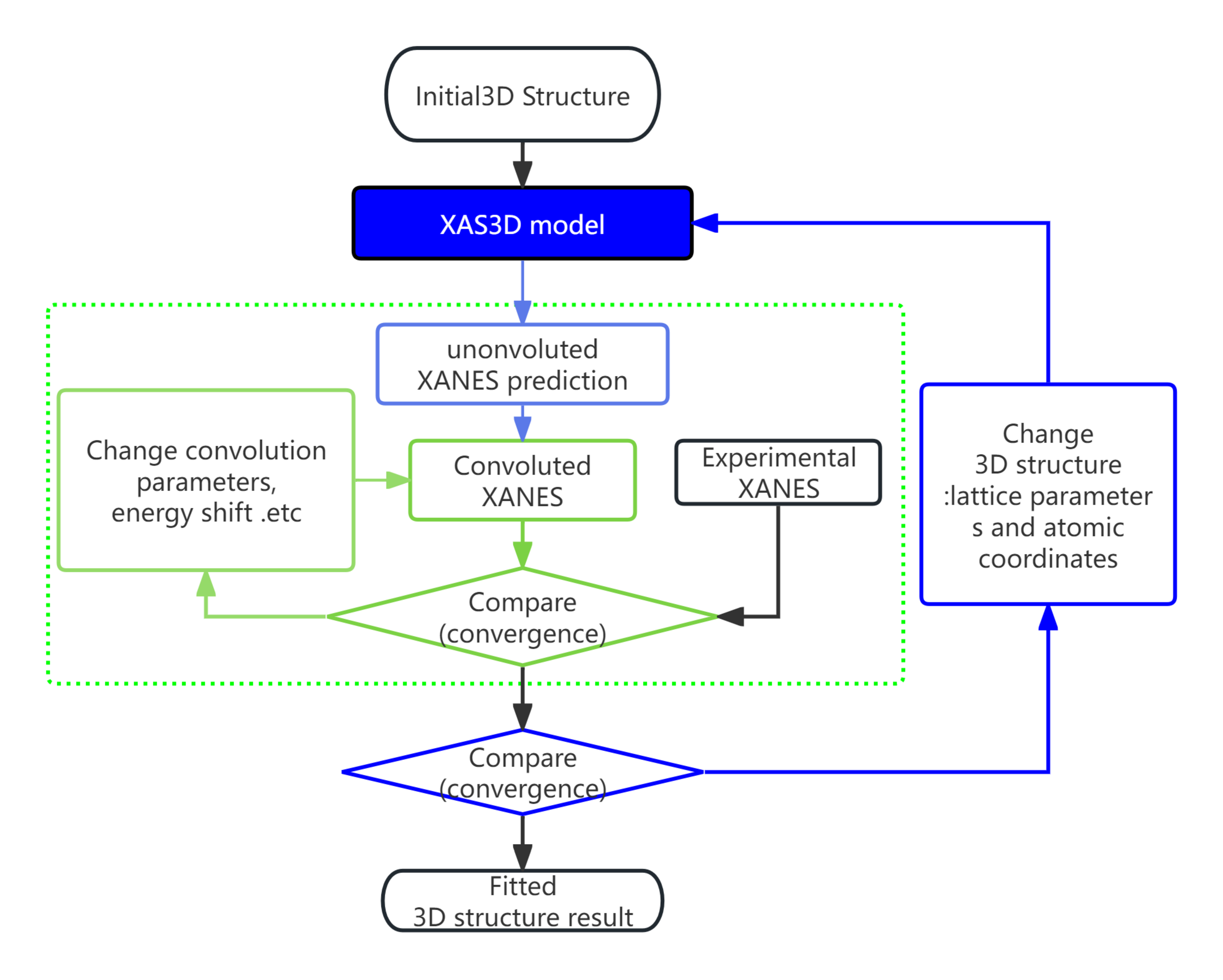}
	\caption[short]{The flowchart of XANES fit combined unconvoluted XANES prediction model and optimization algorithm.}
	\label{fig:unconvFitting}
\end{figure}
\begin{figure}[H]
	\centering
	\includegraphics[width=0.7\linewidth]{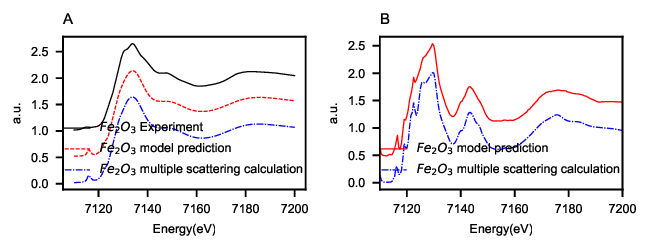}
	\caption[short]{(A)Comparison of the experimental spectra, the model predicted spectra and multiple scattering calculated one based on the fitted 3D structure of $Fe_2O_3$ material.(B)Comparison between the model-predicted unconvoluted spectra and the unconvoluted spectra obtained from multiple scattering calculations.}
	\label{fig:expfitting}
\end{figure}

Beyond the single spectra calculation for a given 3D structure, we also investigate quantitative 3D structure analysis method of material from XANES.
Here, we propose a feasible technical approach based on XANES fitting using the fine-tuned model.
For given material, an initial small dataset is first constructed through structural sampling. 
Through fine-tuning of the element generalization XAS3D model using this dataset, the XAS3D model for the material is derived.
In the modified XANES fitting, this fine-tuned XAS3D model replaces multiple scattering calculation to predict the spectrum for any proposed 3D structure.
By iteratively varying the three-dimensional structure and comparing the predicted spectrum from the XAS3D model with the experimental spectrum, a structure consistent with the experimental data can be determined, thereby achieving the quantitative determination of the material's 3D structure. This fitting process comprises two nested layers: the outer layer corresponds to the aforementioned structural optimization loop, while the inner layer fits non-structural parameters such as broadening, energy shift, and normalization factor as shown in Fig.. Within the inner layer, the XAS3D model predicts the unconvoluted XANES for a given structure. This unconvoluted spectrum is then compared against the experimental spectrum after applying variable broadening, energy shifts, and amplitude scaling to determine the optimal non-structural parameters as shown in Fig.\ref{fig:unconvFitting}.
The feasibility of this technical approach is illustrated through application to the $Fe_2O_3$ material system. Fitting results are shown in Fig.\ref{fig:expfitting}. Here using $Fe_2O_3$ as an example, we performed fitting of its standard XANES spectrum based on the predicted spectrum from the XAS3D model, thereby illustrating a workflow for applying the model to process experimental spectra.

\par

\section{Conclusion} 
To summarize, in this work, we introduce an XANES prediction framework that has demonstrated its efficacy in forecasting XANES spectra for multiple absorption edges of diverse elements.
The predictive capability of the framework extends across both hard X-ray XAS, as demonstrated by the K-edge absorption features of 3d and 4d transition metal elements, and soft X-ray XAS, exemplified through the S K-edge spectrum.
A salient feature of this approach lies in its universality; a solitary model is adept at predicting the XANES responses of myriad elements, obviating the need for element-specific models. This streamlined strategy not only enhances computational efficiency but also underscores the model's versatility.
The proposed framework accepts the intricate three-dimensional architecture of molecules as input data and yields precise XANES predictions as output, thereby showcasing remarkable generalizability. Furthermore, it boasts resilience to fluctuations in hyperparameter settings, ensuring stable performance even amidst variations in these tunable parameters. This inherent robustness translates into a tangible reduction in the computational resources expended during the hyperparameter optimization phase, rendering the framework highly pragmatic and resource-efficient for real-world applications.
This methodology can function as an online data analysis instrument utilized in X-ray absorption spectroscopy (XAS) beamlines pertinent to the disciplines of physics and materials science. It empowers researchers to instantaneously validate their hypothesized 3D
structure and calibrate experimental methodologies grounded on the analysis results. 

\section{Ackonwledgements}
We acknowledge financial support from the Jialin Xie Foundation from Institute of high energy physics, Chinese academy of sciences. The authors acknowledge discussions with Haifeng Zhao and Lirong Zheng.

\end{document}